\documentclass[a4paper,fleqn]{cas-dc}
\usepackage{hyperref}
\usepackage{placeins}
\usepackage[authoryear,longnamesfirst]{natbib}
\usepackage{booktabs}
\usepackage{pifont}
\usepackage{orcidlink}
\usepackage{caption}
\newcommand{\cmark}{\ding{51}}
\newcommand{\xmark}{\ding{55}}

\usepackage{algorithm}
\usepackage{algpseudocode}
\usepackage{amsmath}   
\usepackage{hyperref}
\usepackage{orcidlink}%
\usepackage[table]{xcolor}  
\usepackage{multirow}
\usepackage{algorithm}
\usepackage{graphicx} 
\usepackage{xcolor}
\usepackage{float}  
\usepackage{amsmath}
\usepackage{amsfonts}
\usepackage{textcomp}
\def\BibTeX{{\rm B\kern-.05em{\sc i\kern-.025em b}\kern-.08em
    T\kern-.1667em\lower.7ex\hbox{E}\kern-.125emX}}
\begin{document}
\let\WriteBookmarks\relax
\def\floatpagepagefraction{1}
\def\textpagefraction{.001}

\shorttitle{GWDM: A Guided Wavelet Diffusion Model for Realistic Histopathology Image Generation}

\shortauthors{Abdullah}

\title[mode=title]{GWDM: A Guided Wavelet Diffusion Model for Realistic Histopathology Image Generation}
\author[1]{Abdullah}
\ead{abdullah@my.jcu.edu.au}
\credit{Conceptualization, Methodology, Software, Data analysis, Writing - Original draft, Writing - Review \& editing, Validation}

\author[1]{Tao Huang}
\ead{tao.huang1@jcu.edu.au}
\credit{Writing - Review \& editing, Formatting guidance}

\author[1]{Ickjai Lee}
\ead{ickjai.lee@jcu.edu.au}
\credit{Writing - Review \& editing, Formatting guidance}

\author[2]{Lei Bi}
\ead{lei.bi@sjtu.edu.cn}
\credit{Writing - Review \& editing, Formatting guidance}

\author[1]{Euijoon Ahn}
\ead{euijoon.ahn@jcu.edu.au}
\cormark[1]
\credit{Supervision, Project administration, Writing - Review \& editing}

\affiliation[1]{organization={College of Science and Engineering},
    addressline={James Cook University},
    city={Cairns},
    postcode={QLD 4878},
    country={Australia}.}

\affiliation[2]{organization={Institute of Translational Medicine},
    addressline={Shanghai Jiao Tong University},
    city={Shanghai},
    country={China}.}

\begin{abstract}
\noindent\textbf{Background:} Histopathology images are recognized as a reliable and comprehensive modality for Deep Learning (DL) based diagnosis for a range of medical diseases. However, the effectiveness of DL approaches is constrained by the limited availability of large-scale annotated datasets, as collecting expert-labeled medical data is costly and time-intensive. Recent advances in generative models, particularly diffusion models, have demonstrated strong capabilities in synthesizing medically relevant content and supporting dataset augmentation. Existing generative frameworks treat histopathology patches as generic natural images, failing to enforce biological constraints such as histomorphological structures and heterogeneous staining patterns essential for diagnostically valid synthesis.\\[0.5ex]
\noindent\textbf{Methods:} We propose a guided wavelet diffusion model (GWDM) framework that captures image details at multiple scales using wavelets, and leverages controllable conditioning based on domain-specific histopathological features (e.g., tissue patterns, color variations, etc.). We further formulate a histomorphological guidance loss applied to both real and generated patches. A loss ramping strategy that gradually introduces these constraints during training, ensuring stable optimisation and robust modelling of tissue architecture. \\[0.5ex]
\noindent\textbf{Results:} GWDM was evaluated on three benchmark datasets, including BCSS, MoNuSeg, and PAIP2019, consistently outperforming state-of-the-art generative methods. Ablation studies confirmed that domain-guided conditioning, perceptual loss, and structural similarity loss each contribute meaningfully to synthesis quality. Downstream nuclei and tissue segmentation experiments further validated the clinical utility of GWDM-generated images, demonstrating consistent segmentation performance improvements when synthetic data augments real training sets.\\[0.5ex]
\noindent\textbf{Conclusions:} This work shows that domain-specific conditioning and dual-loss optimization generate realistic histopathology patches, reducing data scarcity and improving clinical downstream segmentation performance.
\end{abstract}





\maketitle

\section{Introduction}
Whole Slide Imaging (WSI) has enabled precise microscopic analysis of tissue samples, aiding in accurate diagnosis, prognosis, and treatment \cite{In12}. However, WSI analysis is typically performed on small fixed-size histopathology patches, which remains challenging due to subtle variations in tissue morphology, complex staining patterns, and the high cost of expert annotation \cite{aydin2026improving}. Deep learning-based Computer-Assisted Diagnosis (CAD) algorithms have been used to address these challenges and they have shown great success in automatic analysis of histopathology \cite{In12, In13}. The performance of these algorithms, however, greatly relies on the availability of large-scale data, which is often difficult and expensive to obtain. The process of acquiring histopathology images includes tissue collection through biopsies or endoscopy, slicing the tissue into thin sections, and mounting them on glass slides. The slides then undergo staining, typically with hematoxylin and eosin (H\&E), for better visualization of cellular structures. Subsequently, high-resolution imaging systems digitize the slides, creating gigapixel-scale WSIs for in-depth analysis \cite{In1}. The intricate nature of this process introduces a great hindrance in the development of reliable computational algorithms.

\begin{figure*}[t]
    \centering
\includegraphics[width=1\textwidth]{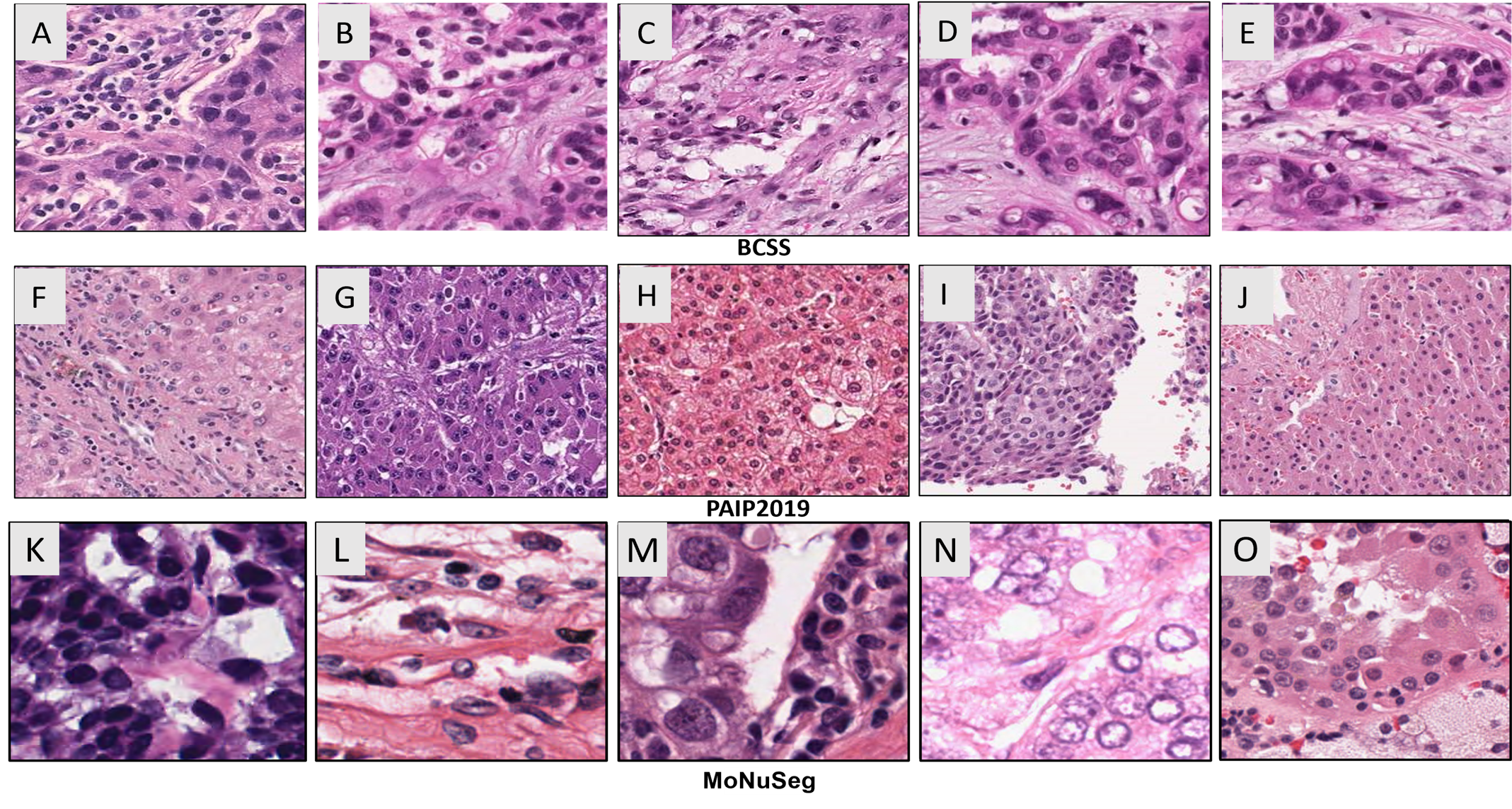}
    \caption{Representative samples from the BCSS, PAIP2019, and MoNuSeg datasets showing diverse histomorphological features. (A) Tightly packed dark purple breast cancer clusters; (B) irregular tumor strands in pink fibrous tissue; (C) densely crowded cells at high magnification; (D) disorganized tumor with mixed purple-pink regions; (E) sheets of packed dark purple cells; (F) irregular pink-stained liver cancer cells of varied size; (G) blue immune cells surrounding liver tumor; (H) purple cells forming gland-like structures; (I) dense, chaotic cell clusters; (J) scattered cells in loose pink tissue; (K) high-density regions with pleomorphic nuclei; (L) spindle-cell stromal architecture; (M) glandular formations with nuclear palisading; (N) adipose-glandular mixed tissue; and (O) inflammatory mixed cells.}
    \label{fig:fig1}
\end{figure*}

As a result, developing methods that generate realistic histopathology has become a focus area of research in recent years. Current state-of-the-art (SOTA) techniques for histopathology image generation predominantly leverage deep learning-based approaches such as Variational Autoencoders (VAEs) \cite{In3} and Generative Adversarial Networks (GANs) \cite{In4}. These methods have demonstrated remarkable success in various domains of digital pathology. Despite these advancements, generating high-resolution histopathology patch image with fine structural details remains challenging due to issues such as model instability, and low model convergence \cite{mescheder2018training}. Recently, Denoising Diffusion Probabilistic Models (DDPMs) \cite{In14} have demonstrated superior generative quality when compared to earlier approaches such as VAEs and GANs. DDPMs progressively add Gaussian noise to an input image over multiple time steps until it becomes fully noisy, then learn to reconstruct the original image from this noisy state. These models have been successfully applied to generate realistic medical images. For instance, Franzes et al. \cite{In16} demonstrated its superior performance over GANs across various medical imaging applications including radiology, ophthalmology, and histopathology.
Furthermore, Shen et al. \cite{In10} also developed StainDiff, a method that performs stain style transfer by repeatedly applying the diffusion process at varying noise levels. Since each diffusion process gives a slightly different result, a self-ensemble is formed by averaging the outputs, improving stability and quality. DDPMs excel in generating high-quality images with fine details, however, they still suffer from slow inference speeds, especially for generating high-resolution images. This trade-off between image quality and generation time remains a key challenge. To address this limitation, a hybrid approach called Denoising Diffusion GAN (DDGAN) \cite{In5} was introduced, which combines the strengths of diffusion models and GANs. This greatly reduces the number of denoising steps required during inference time. 

Recently, wavelet transforms are incorporated into the diffusion process to lower the data dimensionality, which leads to the development of the Wavelet-Diffusion Model (WDM) \cite{In17}. This approach reduces inference time by approximately fourfold compared to traditional DDPMs. This combination maintained generative quality while significantly reducing inference time. The wavelet transform decomposes an image into multi-frequency components, preserving spatial correspondence and directional properties. It effectively extracts features and structures by separating image content into frequency bands at various scales, improving feature extraction \cite{In18}. Following this, Friedrich et al. \cite{In19} applied WDM for high-resolution medical image synthesis, reducing memory usage by transforming images into the wavelet frequency domain and enabling spatial dimensionality reduction. This approach is particularly beneficial for generating high-resolution images, such as 3D histopathology gigapixel images. However, WDM has inherent limitations, performing decomposition beyond the first level increases the likelihood of losing critical information. Although deeper decomposition can improve computational speed, it often results in a decline in image quality. WDM \cite{In17}, built upon a GAN backbone \cite{In4}, leverages the generative strength of GANs during both training and inference. However, its conventional approach of denoising Gaussian noise without integrating domain-specific knowledge can lead to a loss of structural and semantic consistency. This is an important gap, where preserving structural and semantic consistency is critical for accurate analysis. As such, in this paper, we introduce a Guided Wavelet Diffusion Model (GWDM), which incorporates histomorphological features to guide the denoising process. This guidance ensures consistency between real and generated patches, preserving clinically relevant tissue structure (see Fig. \ref{fig:fig1}). GWDM is designed specifically for histopathology patch generation to support reliable downstream clinical tasks. Our main contributions are summarized as follows:

\begin{itemize}
\item[1)] We propose a GWDM for generating high-resolution histopathology images by conditioning the diffusion process on measurements \(y\) that captures morphological (e.g., nuclei density, edges, texture), multi-scale spatial, and staining characteristics. This integration of domain-specific features enhances image fidelity and downstream analytical performance, enabling robust modeling of tissue architecture without requiring paired data or spatial annotations.

\item[2)] We formulate a histomorphological guidance loss comprising three domain-specific measurement terms extracted through a measurement operator $\Psi$ applied to both real and generated patches. A loss ramping function $\rho(\epsilon)$ gradually introduces these constraints during training, ensuring stable optimization and structural consistency at both local and global scales.

\item[3)] We validate the proposed model on three benchmark histopathology datasets and show that GWDM achieves higher synthesis quality than existing generative methods. Ablation studies and a downstream segmentation task further confirm the effectiveness of the histomorphological guidance and the clinical utility of the synthesized patches.

\end{itemize}
The remainder of this manuscript is organized as follows. Section II reviews related work. Section III presents a comprehensive description of the proposed method. Section IV outlines the experimental setup. Section V reports the results, and Section VI presents the downstream task. Section VII discusses the findings, and Section VIII concludes the study.

\section{RELATED WORK}

\subsection{Diffusion Model in Medical Imaging}
Diffusion models have recently shown promising results in medical imaging by generating high-quality images, allowing better control during image synthesis and providing more reliable uncertainty estimation compared to GAN-based methods\cite{kazerouni2022diffusion}. For example, Xu et al. \cite{a19} utilized wavelet-based optimization within a diffusion framework for sparse-view CT reconstruction, addressing the ill-posed nature of incomplete data recovery. Similarly, Chen et al. \cite{li2023ddmm} proposed DDMM-Synth, leveraging denoising diffusion models for cross-modal image synthesis, effectively refining missing information and outperforming SOTA supervised methods under noisy or limited-projection conditions. Özbey et al. \cite{ozbey2023unsupervised} extended diffusion-based synthesis to unsupervised cross-modal MRI translation, demonstrating that score-based models can bridge modality gaps without paired training data. Kim et al. \cite{lee2025lesion} proposed a lesion-aware post-training framework incorporating pixel-space objectives into pre-trained LDMs for medical image-to-image translation, demonstrating improved lesion delineation and overall image quality for CT-to-MRI synthesis in acute ischemic stroke patients. In dermatology, Farooq et al. \cite{farooq2024derm} employed stable diffusion models to generate high-quality and diverse synthetic dermoscopic images, enhancing dataset variety and addressing the scarcity of labeled skin lesion data. Furthermore, Akrout et al. \cite{akrout2023diffusion} showed that diffusion-generated synthetic skin images enhance the accuracy of classifier when they were combined with real data.

\subsection{Diffusion-Based Histopathology Image Generation}
Diffusion-based generative modeling has recently gained significant attention in computational pathology. Yellapragada et al.~\cite{a20} introduced PathLDM, which combines textual descriptions with a latent diffusion model (LDM)~\cite{a21} through a two-stage encoding and diffusion pipeline for histopathology image synthesis. Similarly, Nakhli et al.~\cite{shrivastava2023nasdm} proposed NASDM, which generates realistic $H\&E$ tissue images from semantic instance masks of up to six nuclei types, improving controllability for nuclei segmentation augmentation. More recently, Bhosale et al.~\cite{bhosale2025pathdiff} introduced PathDiff, a diffusion-based framework that conditions generation on unpaired masks and diagnostic text, improving tissue structure control, text-image alignment, and downstream nuclei segmentation performance. Moghadam et al.~\cite{In9} used a diffusion-based framework focused on morphological patterns and colour normalization to generate realistic brain cancer histopathology images. Xu et al.~\cite{a22} introduced a hybrid diffusion architecture combining vision transformers and diffusion autoencoders to better capture complex structural and textural features. Similarly, Porter et al.~\cite{a23} developed conditional diffusion models with a dual-weighting mechanism that preserves both perceptual tissue structures and subtle statistical characteristics, improving the realism of synthesised histopathology images.


\subsection{Domain-Guided Generative Conditioning}
Most diffusion models start from pure Gaussian noise, leading to slow convergence and generic outputs \cite{a15, nichol2021improved}. However, by introducing conditioning mechanisms address these limitations by steering generation toward domain-specific targets. For instance, Zhang et al. \cite{zhang2023adding} introduced ControlNet, which incorporates spatial conditioning signals into pretrained diffusion models. It enables domain-specific generation without disrupting the pretrained backbone. Similarly, ProtoDiffusion \cite{ho2022classifier} employs class prototypes as conditioning inputs, which provides structured guidance to accelerates convergence. Furthermore, Chambon et al. \cite{chambon2022roentgen} proposed presented a method leveraging radiology reports as conditions for chest X-ray image synthesis. In parallel, SegGuidedDiff \cite{konz2024anatomically} guides brain MRI generation using anatomical segmentation masks at each sampling step, improving the anatomical fidelity of the synthesized outputs. Moreover, Moreover, Moghadam et al. \cite{In9} further extended this idea to histopathology image generation, where genotype data were used as guidance to enable morphologically and genomically diverse synthesis. These works collectively motivate the integration of domain knowledge into diffusion models, as adopted in GWDM.

\begin{figure*}[t]
    \centering
    \includegraphics[width=1\textwidth]{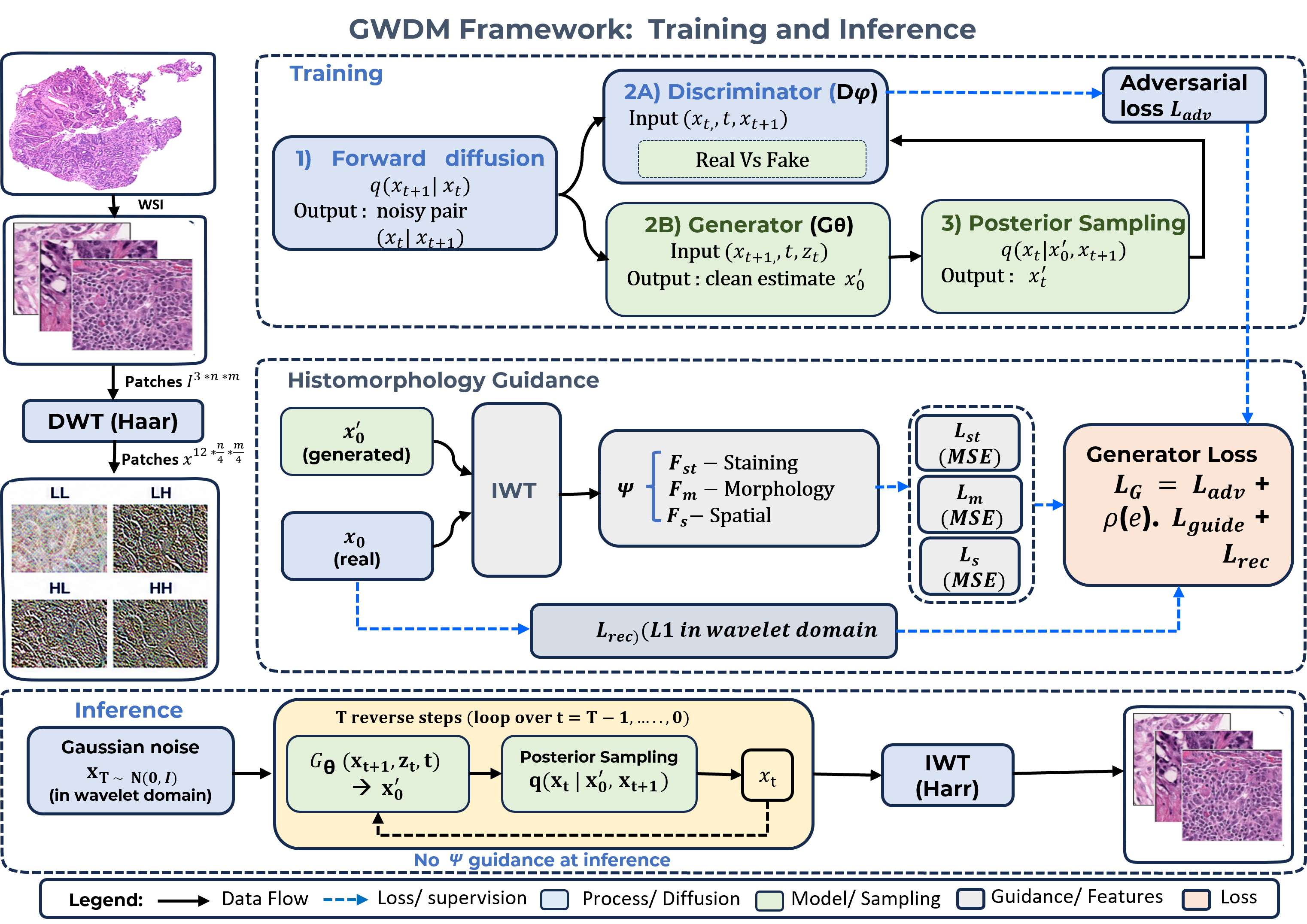}
    \caption{Architecture of the proposed GWDM framework for training and inference. Real image patches are first decomposed using Haar-based discrete wavelet transformation (DWT) into multi-channel wavelet coefficients. During training, forward diffusion generates noisy image transitions that are processed by a generator–discriminator framework and posterior sampling. Histomorphology guidance is incorporated through a measurement operator that extracts staining, morphological, and spatial features, producing corresponding loss terms. These guidance losses are combined with adversarial and reconstruction losses for optimization. During inference, Gaussian noise in the wavelet domain undergoes iterative reverse diffusion, followed by rescaling and inverse wavelet transformation (IWT), to generate synthesized image patches.}
    \label{fig:block_diagram}
\end{figure*}

\section{Method}
\subsection{Problem Formulation}
\label{sec:problem_formulation}
Suppose $I_0$ denotes a histopathological image patch used in downstream tasks such as cancer diagnosis, tissue segmentation, and classification. We first extract domain-relevant features from $I_0$ that capture key aspects such as staining variability, morphological detail, and spatial structure. This is formalized as
\begin{equation}
\Psi(I_0) = \left\{\Psi_{\mathrm{stain}}(I_0),\, \Psi_{\mathrm{morph}}(I_0),\, \Psi_{\mathrm{spatial}}(I_0)\right\},
\label{eq:psi_decomposition}
\end{equation}
where $\Psi_{\mathrm{stain}}$, $\Psi_{\mathrm{morph}}$, and $\Psi_{\mathrm{spatial}}$ denote stain color ~\cite{ruifrok2001quantification}, morphological \cite{In9,a27}, and spatial descriptors \cite{schmitz2021multi}, respectively. Here, $\Psi$ acts as a domain-informed measurement operator applied to $I_0$, yielding a compact conditioning representation $y$, generally expressed as
\begin{equation}
y = \Psi(I_0).
\label{eq:measurement}
\end{equation}
We aim to synthesize high-fidelity patch images $\hat{I}_0$ whose histomorphological measurements match those of real patches, i.e., $\Psi(\hat{I}_0) \approx \Psi(I_0)$, while learning a GAN-based wavelet diffusion generator  $G_\theta$ within GWDM. Our objective is to preserve structural integrity, staining fidelity, and spatial consistency in the synthesized images. Furthermore, we will validate the clinical utility of the synthesized patches by performing a downstream segmentation task.

\subsection{Proposed Method}
\label{sec:proposed_method}
GWDM synthesizes high-resolution histopathology patches $\hat{I}_0$ whose domain measurements are constrained to match those of real tissue. The model unifies wavelet-domain diffusion with a GAN-based score learner and enforces histomorphological guidance through the measurement operator $\Psi$ defined in Eqs.~\eqref{eq:psi_decomposition}-\eqref{eq:measurement}, steering synthesis toward biologically faithful structure.
\subsubsection{Wavelet Representation}
\label{sec:wavelet}
A clean input image is represented as $I_0 \in \mathbb{R}^{3 \times n \times m}$, where $n$ and $m$ denote the spatial dimensions. We apply a discrete wavelet transform (DWT) to obtain low and high-frequency subbands, which are concatenated into a multi-channel wavelet representation $x_0 \in \mathbb{R}^{12 \times \frac{n}{2} \times \frac{m}{2}}$, reducing the spatial resolution by a factor of four. Diffusion and adversarial training are performed in this transformed domain. 

 
\subsubsection{Training Stage}
The forward diffusion process \(q(x_{t}|x_0)\) is applied to the transformed image, where \(q\) denotes a conditional distribution, \(x_0\) is the clean wavelet coefficients, and \(x_{t}\) represents the progressively noised state at time step $t$. The forward diffusion process is defined as

\begin{equation}
q(x_{t} \mid x_0) = \mathcal{N}\left(x_{t}; \sqrt{\alpha_t} \, x_0, (1-\alpha_t )\mathbf{I} \right),
\label{eq:1}
\vspace{-5pt}
\end{equation}
where $\mathcal{N}(\cdot)$ denotes the Gaussian distribution, $\alpha_t$ 
represents the variance schedule controlling the amount of noise added at each step, constrained by $0 \leq \alpha_t \leq 1$, and $\mathbf{I}$ is the identity matrix. During training, we sample a timestep $t$ and construct a consecutive noisy pair $(x_t, x_{t+1})$ from the real wavelet coefficients $x_0$. The discriminator is trained to distinguish real denoising transitions from generator-produced transitions, conditioned on the noisier state $x_{t+1}$, providing a discriminative signal that guides model optimization. The generator $G_\theta$ samples latent noise $z_t \sim \mathcal{N}(0, \mathbf{I})$ and maps the noisy state $x_{t+1}$ to produce an estimated $\hat{x}_0$ of a clean image as shown in Fig. \ref{fig:block_diagram}.
\begin{equation}
\hat{x}_0 = G_\theta(x_{t+1}, z_t, t).
\label{eq:generator}
\end{equation}
The estimated clean coefficients $\hat{x}_0$ are used along two branches. First, the intermediate state for adversarial learning is sampled from the analytic DDPM posterior $q(x_t \mid \hat{x}_0, x_{t+1})$. Second, for histomorphological guidance, $\hat{x}_0$ is reconstructed to RGB by the inverse wavelet transform, $\hat{I}_0 = \mathcal{W}^{-1}(\hat{x}_0)$, and the corresponding measurement is computed as
\begin{equation}
\hat{y} = \Psi(\hat{I}_0),
\label{eq:fake_measurement}
\end{equation}
where $\hat{y}$ is the measurement of the generated image and is compared with the reference measurement $y = \Psi(I_0)$ from Eq.~\eqref{eq:measurement}. Treating $y$ as the observation, the posterior over the generated image follows from Bayes' theorem,
\begin{equation}
p(\hat{I}_0 \mid y) = \frac{p(y \mid \hat{I}_0)\, p(\hat{I}_0)}{p(y)},
\label{eq:bayes}
\end{equation}
and taking gradients of the log-posterior with respect to $\hat{I}_0$ gives
\begin{equation}
\nabla_{\hat{I}_0} \log p(\hat{I}_0 \mid y) 
= \nabla_{\hat{I}_0} \log p(y \mid \hat{I}_0) 
+ \nabla_{\hat{I}_0} \log p(\hat{I}_0).
\label{eq:logpost}
\end{equation}
Assuming a Gaussian likelihood, the data term reduces to the squared 
measurement discrepancy,
\begin{equation}
-\log p(y \mid \hat{I}_0) \propto \lVert \hat{y} - y \rVert_2^2 
= \lVert \Psi(\hat{I}_0) - \Psi(I_0) \rVert_2^2,
\label{eq:gauss}
\end{equation}

so the likelihood term in Eq.~\eqref{eq:logpost} is implemented as training time feature matching rather than as an explicit sampling time gradient. Specifically, we minimize mean squared errors between the stain, morphology, and spatial components of $\hat{y}$ and $y$, together with an adversarial denoising loss and an wavelet-domain reconstruction loss shown in Algorithm~\ref{alg:train_step}.

\subsubsection{Measurement Model}
\label{Measurement model}
To achieve precise and consistent image synthesis, we propose a conditional measurement model that jointly incorporates morphological, spatial, and staining features. This model can be formalized as: 
\begin{equation} \label{eq:10}
    y = \Psi(I_0),
\end{equation}
where the measurement operator is defined as:
\begin{equation}
    \Psi = 
    \begin{cases}
        F_m, & \text{morphological features,} \\
        F_s, & \text{spatial features,} \\
        F_{st}, & \text{staining features.}
\end{cases}
\label{eq:transfer_operator}
\end{equation}
Here $F_m$ reprsents morphological features ~\cite{moghadam2023morphology},  $F_s$ captures spatial features ~\cite{rani2025efficient}, and $F_{st}$ corresponds to staining features \cite{vaishnani2023impact}. The morphological features $F_m$ encompass nuclear density, edges of cellular and tissue structures, and texture patterns that characterize the microscopic morphology. The spatial features $F_s$ are extracted using multi-scale Gaussian kernels, capturing information across varying spatial resolutions. The staining features $F_{st}$ are derived from H\&E color and optical-density characteristics, representing the staining intensity and color variations relevant to tissue appearance. 

\subsubsection{Histomorphological Guidance Losses}
\label{sec:perceptual_structural_losses}
Histomorphological guidance aligns the generated patch with its real counterpart in clinically meaningful appearance by comparing features extracted from the generated image $\hat{I}_0$ with those of the corresponding real patch $I_0$. For each training pair, the reference features $y$ are computed from the real patch and kept fixed, while gradients are propagated only through the generated branch. The guidance objective is combined with the adversarial denoising loss into a single generator objective:
\begin{equation}
\mathcal{L}_G = \mathcal{L}_{\mathrm{adv}}
  + \rho(e)\,\bigl\|\Psi(\hat{I}_0) - y\bigr\|_2^2
  + \omega(e)\,\bigl\|\hat{x}_0 - x_0\bigr\|_1,
\label{eq:objective}
\end{equation}
where $\mathcal{L}_{\mathrm{adv}}$ is the non-saturating adversarial loss on denoising transitions, and $\rho(e)$, $\omega(e)$ are epoch-dependent schedules: $\rho(e)$ linearly ramps up the guidance term during early training, while $\omega(e)$ attenuates the stabilizer, so guidance dominates as training progresses. 

\begin{algorithm}[t]
\caption{Guided Wavelet Diffusion Model Training}
\label{alg:train_step}
\begin{algorithmic}[1]
\Require Real image $I$; wavelet transform $\mathcal{W}(\cdot)$; $\Psi(\cdot)=\{F_{\mathrm{st}}, F_{\mathrm{m}}, F_{\mathrm{s}}\}$;  discriminator$D_\phi$; generator $G_\theta$; schedules $\rho(e)$; learning rate $\eta$

\State $x_0 \gets \tfrac{1}{2}\,\mathcal{W}(I)$
\State Sample $t \sim \text{Uniform}\{0, \ldots, T-1\}$, 
       $z \sim \mathcal{N}(0, \mathbf{I})$

\State $(x_t, x_{t+1}) \gets q(x_0, t)$
\State $\hat{x}_0 \gets G_\theta(x_{t+1}, z, t)$
\State $\tilde{x}_t \gets \mathrm{Posterior}\hat(x_0, x_{t+1}, t)$
\State $\mathcal{L}_{\mathrm{adv}} \gets \mathcal{L}_{\mathrm{adv}}(\tilde{x}_t, x_{t+1}, t)$
\State $\hat{I}_0 \gets \mathcal{W}^{-1}(\hat{x}_0)$
\State Compute $y \gets \Psi(I_0)$ \Comment{fixed reference features}
\State $\mathcal{L}_G \gets \mathcal{L}_{\mathrm{adv}}
  + \rho(e)\,\bigl\|\Psi(\hat{I}_0) - y\bigr\|_2^2
  + \omega(e)\,\bigl\|\hat{x}_0 - x_0\bigr\|_1$
\State $\theta \gets \theta - \eta\,\nabla_\theta \mathcal{L}_G$
\end{algorithmic}
\end{algorithm}


\subsubsection{Inference Stage}
\label{sec:inference}

During inference, the trained generator $G_\theta$ synthesizes histopathology patches by progressively denoising random Gaussian noise in the wavelet domain. Sampling is initialized as
\begin{equation}
x_T \sim \mathcal{N}(0, \mathbf{I}),
\label{eq:inference_init}
\end{equation}
where $x_T$ has the same dimensionality as the wavelet representation $x_0$. The reverse process iterates for $t = T-1,\ldots,0$, the generator predicts a clean wavelet estimate
\begin{equation}
\hat{x}_0 = G_\theta(x_{t+1}, z_t, t),
\label{eq:inference_generator}
\end{equation}
where $z_t \sim \mathcal{N}(0,\mathbf{I})$ is latent noise and $x_{t+1}$ denotes the current noisy state. The previous state is then obtained by posterior sampling,
\begin{equation}
x_t \sim q(x_t \mid \hat{x}_0, x_{t+1}),
\label{eq:inference_posterior}
\end{equation}
which incorporates controlled stochasticity through the analytic DDPM posterior. The synthesized wavelet representation is then rescaled and mapped back to RGB  via the inverse wavelet transform $\mathcal{W}^{-1}$, yielding the final three-channel histopathology image (see Fig.~\ref{fig:block_diagram}). No histomorphological guidance term is applied at inference, image quality rests entirely on the generator learned under the adversarial and measurement-consistency objectives of Eq.~\eqref{eq:objective}.

\section{EXPERIMENTS}
\begin{table}[]
\centering
\caption{Key hyperparameters impacting GWDM performance.}
\begin{tabular}{ll}
\hline
\textbf{Hyperparameter} & \textbf{Value} \\
\hline
Learning Rate (G / D) & $1.6\times10^{-4}$ / $1.25\times10^{-4}$ \\
Batch Size & 8 \\
Number of Diffusion Timesteps & 2 \\
EMA Decay ($\tau$) & 0.9999 \\
Number of Residual Blocks & 2 \\
Attention Resolution & 16 \\
Linearly Ramp $\rho(e)$ & 50\\
Optimizer & Adam \\
\hline
\end{tabular}
\label{tab:key_hyperparams}
\end{table}
\subsection{Datasets and Preprocessing} The proposed method was rigorously evaluated using three benchmark histopathology image datasets: A multi-organ nucleus segmentation challenge (MoNuSeg) \cite{kumar2019multi}, the Breast Cancer Semantic Segmentation (BCSS) \cite{A9}, and the Pathology Artificial Intelligence Platform (PAIP) 2019 challenge dataset \cite{A8}. These datasets were utilized to quantitatively and qualitatively assess the model’s efficiency in generating high-fidelity histopathology images. 

\subsubsection{BCSS dataset} 
The BCSS dataset \cite{A9} comprises H\&E-stained WSIs from breast cancer cases. Regions of interest (ROIs) were initially annotated by 25 participants and subsequently validated by a senior pathologist to ensure accuracy. Each ROI averages 1.18~mm$^{2}$ (SD: 0.80~mm$^{2}$) at 40$\times$ magnification (0.25~\textmu m/px). 
As part of preprocessing, WSIs were divided into non-overlapping 256$\times$256 pixel patches at 40× magnification, ensuring consistent spatial resolution for analysis. 

\subsubsection{MoNuSeg dataset} 
The MoNuSeg dataset~\cite{kumar2019multi} comprises 37 histopathology images of size $1000 \times 1000$ pixels, containing 21,623 expert-annotated nuclear boundaries across  seven organ types (breast, liver, kidney, prostate, bladder, colon, and stomach). Images were acquired at $40\times$ magnification from individual patients through TCGA \cite{wang2016practical}, covering both benign and malignant tissue to ensure morphological diversity. For training and testing, each image was partitioned into $256 \times 256$ pixel patches.

\subsubsection{PAIP2019 dataset} 
The PAIP 2019 dataset \cite{A8}, developed as part of the Liver Cancer Segmentation Challenge, provides a benchmark for evaluating both established and novel algorithms in automated liver cancer detection using WSIs. The dataset consists of H\&E-stained WSIs scanned using an Aperio AT2 scanner at 20$\times$ magnification (0.5021~$\mu$m/pixel), with image resolutions ranging from 35,855\,$\times$\,39,407 to 64,768\,$\times$\,47,009 pixels.
For analysis, WSIs were partitioned into non-overlapping 256$\times$256 pixel patches at same magnification, preserving consistent tissue detail.
\begin{table*}[t]
\centering
\caption{Quantitative results of GWDM on the BCSS, MoNuSeg and PAIP2019 datasets, compared against SOTA methods. The best-performing result per metric is in bold.}
\renewcommand{\arraystretch}{1.5}
\begin{tabular}{l l c c c c}
\hline
Datasets & Methods & FID $\downarrow$ & KID $\downarrow$ & pFID $\downarrow$ & Path-FID $\downarrow$ \\
\hline
\multirow{7}{*}{BCSS}
& LDM \cite{a21}                    & 165.33 & 0.186 $\pm$ 0.053 & 58.126  & 1457.785 \\
& Imp-DDPM \cite{nichol2021improved}& 159.21 & 0.148 $\pm$ 0.180 & 140.921 & 1430.780 \\
& DDIM \cite{song2020denoising}     & 145.02 & 0.127 $\pm$ 0.283 & 123.573 & 1431.369 \\
& DD-GAN \cite{In5}                 & 91.34  & 0.075 $\pm$ 0.088 & 88.968  & 1444.288 \\
& SPAD \cite{park2019semantic}      & 90.74  & 0.087 $\pm$ 0.109 & 61.547  & 1089.607 \\
& WDM \cite{In17}                   & 111.91 & 0.101 $\pm$ 0.001 & 86.168  & 1517.665 \\
& GWDM (Ours)                       & \textbf{12.64} & \textbf{0.007 $\pm$ 0.144} & \textbf{27.033} & \textbf{246.881} \\
\hline
\multirow{7}{*}{MoNuSeg}
& LDM \cite{a21}                    & 153.23 & 0.137 $\pm$ 0.015 & 69.889  & 1372.172 \\
& Imp-DDPM \cite{nichol2021improved}& 98.09  & 0.067 $\pm$ 0.008 & 99.117  & 1037.808 \\
& SPAD \cite{park2019semantic}      & 59.38  & 0.049 $\pm$ 0.082 & 55.133  & \textbf{624.464} \\
& DD-GAN \cite{In5}                 & 59.09  & 0.030 $\pm$ 0.179 & \textbf{50.373} & 994.985 \\
& DDIM \cite{song2020denoising}     & 55.35   & 0.032 $\pm$ 0.210 & 56.119  & 651.422 \\
& WDM \cite{In17}                   & 48.25  &  0.022$\pm$ 0.051 & 54.111  & 839.160 \\
& GWDM (Ours)                       & \textbf{45.02} & \textbf{0.020 $\pm$ 0.001} & \textbf{51.589} & 857.123 \\
\hline
\multirow{4}{*}{PAIP}
& LDM \cite{a21}                    & 209.89 & 0.254 $\pm$ 0.001 & 140.167 & 1530.804 \\
& WDM \cite{In17}                   & 19.46  & 0.012 $\pm$ 0.056 & 33.82   & \textbf{240.527} \\
& GWDM (Ours)                       & \textbf{15.76}  & \textbf{0.005 $\pm$ 0.070} & \textbf{28.696}  & 323.360 \\
\hline
\end{tabular}
\label{tab:table_1}
\end{table*}

\subsection{Implementation details}
The proposed framework was developed using PyTorch and equipped with a three-NVIDIA RTX A5500 GPU cluster for model training across each dataset, while inference and evaluation were performed using a single GPU. 
For all datasets, we employed two sampling steps, with image resolutions fixed at 256×256 pixels during both training and testing phases, and a batch size of 8 consistent with another similar study in \cite{In17}. The network parameters, including the learning rate, optimizer configuration, batch size, number of residual blocks, attention resolution, and sampling steps, were adopted from the original WDM architecture to maintain consistency in training dynamics and model performance (see Table \ref{tab:key_hyperparams}).

\subsection{Evaluation metrics}
We evaluate our method using both image-level quality metrics and a 
downstream task that reflects clinical utility. For image-level quality, 
we employ four metrics: Fréchet Inception Distance (FID), Kernel Inception 
Distance (KID), patch-FID (pFID), and path-FID. FID~\cite{a12} measures the similarity between the distributions of real 
images \(I\) and generated images \(\hat{I}\) by comparing the means and 
covariances of their feature representations extracted from a pretrained 
Inception network \cite{szegedy2017inception}:
{\small
\begin{equation}
\text{FID}(I, \hat{I}) = \|\mu_I - \mu_{\hat{I}}\|^2 + 
\mathrm{Tr}\Big(\Sigma_I + \Sigma_{\hat{I}} - 2(\Sigma_I \Sigma_{\hat{I}})^{1/2}\Big),
\label{eq:fid}
\end{equation}}

where \(\mu_I, \Sigma_I\) and \(\mu_{\hat{I}}, \Sigma_{\hat{I}}\) denote the 
means and covariances of the real and generated image features, respectively.  Lower FID values indicate better alignment with the real data distribution and higher generative quality.

KID~\cite{a26} measures distribution similarity using the squared Maximum 
Mean Discrepancy (MMD) between Inception features with a polynomial kernel 
\(k\), providing an unbiased estimator that is more reliable for smaller 
sample sizes:
{\small
\begin{equation}
\text{KID} = \mathbb{E}[k(u, u')] + \mathbb{E}[k(v, v')] - 2\,\mathbb{E}[k(u, v)],
\end{equation}}
where \(u, u'\) are features from real images \(I\), and \(v, v'\) are features from generated images \(\hat{I}\). Lower KID values indicate closer similarity to the real image distribution.




pFID~\cite{yin2025efficient} extends FID by computing the score over local image patches rather than whole images, thereby emphasizing local texture 
and fine-grained structural fidelity. path-FID adapts the FID formulation to 
the histopathology domain by replacing the ImageNet-pretrained Inception 
network with UNI~\cite{chen2024towards}, a pathology foundation model, as the feature  extractor, yielding a domain-specific measure of generative fidelity that is more sensitive to histopathological structure than Inception-based scores. For all four metrics, lower values indicate better agreement with the real data distribution.


\section{Results}
\begin{table}[t]
\centering
\caption{Ablation study of proposed guidance components on the MoNuSeg dataset.
\cmark\ indicates the component is included; \xmark\ indicates it is excluded}
\label{tab:monuseg_ablation}
\resizebox{\columnwidth}{!}{%
\begin{tabular}{lccccc c}
\toprule
\textbf{Variant}
  & $\mathcal{L}_{\mathrm{stain}}$
  & $\mathcal{L}_{\mathrm{morph}}$
  & $\mathcal{L}_{\mathrm{spat}}$
  & $\mathcal{L}_{\mathrm{rec}}$
  & \textbf{Ramp}
  & \textbf{FID}$\downarrow$ \\
\midrule
Baseline (WDM)                         & \xmark & \xmark & \xmark & \cmark & \xmark & 51.49 \\
\midrule
$+$ $\mathcal{L}_{\mathrm{stain}}$ only  & \cmark & \xmark & \xmark & \cmark & \cmark & 49.80  \\
$+$ $\mathcal{L}_{\mathrm{spat}}$ only  & \xmark & \cmark & \xmark & \cmark & \cmark & 48.01 \\
$+$ $\mathcal{L}_{\mathrm{morph}}$ only  & \xmark & \cmark & \xmark & \cmark & \cmark & 47.20 \\
$+$ all losses (no ramp)                 & \cmark & \cmark & \cmark & \cmark & \xmark & 46.42 \\
\textbf{Full model (GWDM, ours)}         & \cmark & \cmark & \cmark & \cmark & \cmark & \textbf{45.02} \\
\bottomrule
\end{tabular}%
}
\end{table}
{\begin{figure*}[t]
    \centering
    \includegraphics[width=1\textwidth]{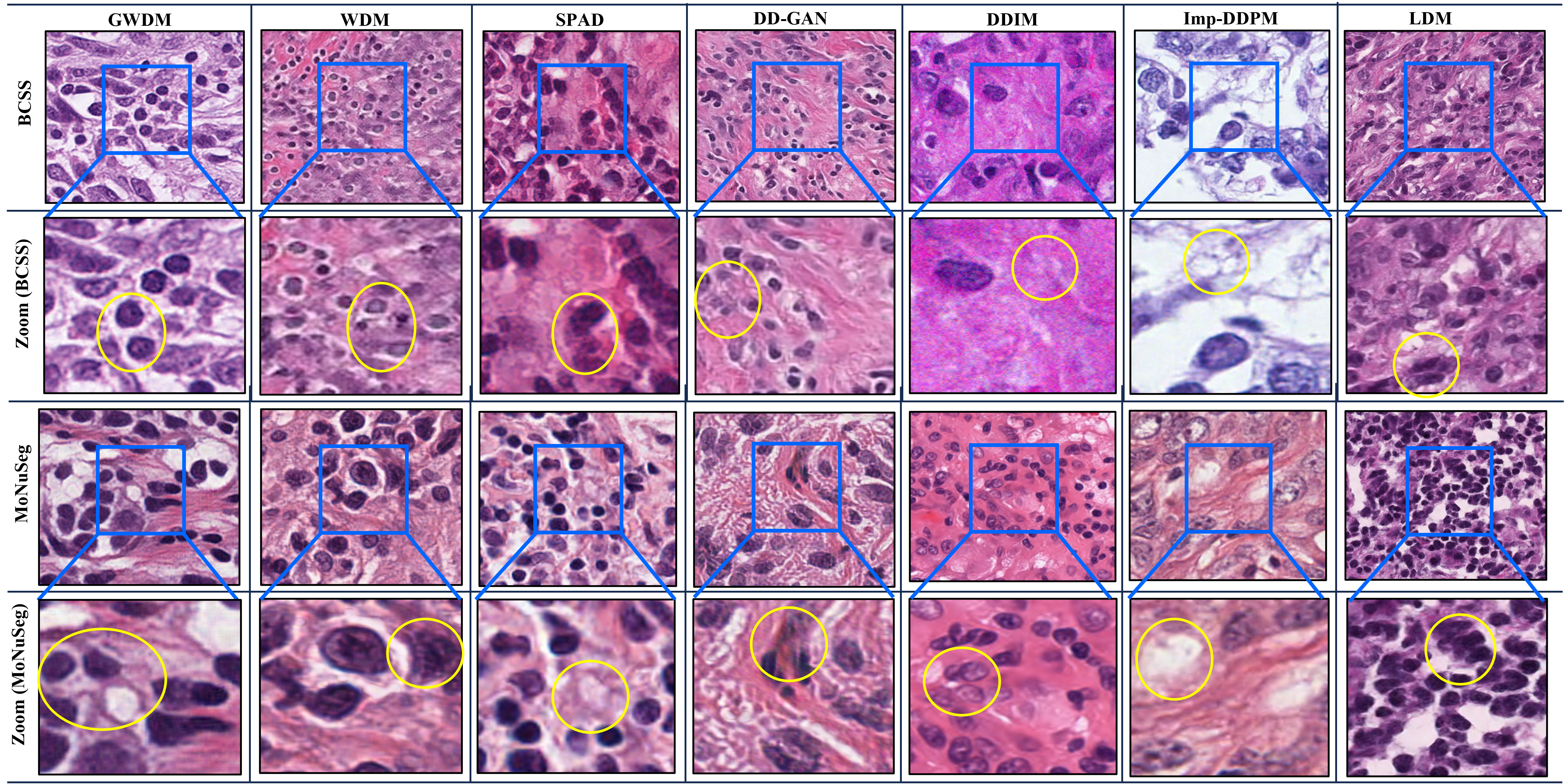}
    \caption{Comparative qualitative analysis of seven generative histopathology models on the BCSS and MoNuSeg datasets. Yellow circles highlight regions where our method preserves texture and more accurate cell boundaries than competing methods. The ground truth visual quality images are shown (see Fig.~\ref{fig:fig1}).}
    \label{fig:GWDM synthetic samples}
\end{figure*}
\subsection{Performance Comparison}
The effectiveness of the proposed GWDM was evaluated by comparing it against several strong generative baselines: LDM \cite{a21}, Imp-DDPM \cite{nichol2021improved}, DDIM \cite{song2020denoising}, DD-GAN \cite{In5}, 
SPAD \cite{park2019semantic}, and WDM \cite{In17}, as shown in Table~\ref{tab:table_1}. Our results show that GWDM consistently outperformed existing models across all datasets, achieving strong fidelity, diversity, and alignment with real data distributions.

On the BCSS dataset, GWDM reached an FID of 12.64, representing a reduction of approximately 86\% compared to both SPAD (90.74) and DD-GAN (91.34), and up to 92\% compared to LDM (165.33). This highlights its strong ability to generate samples that closely match the real data distribution. A similarly strong trend was observed for KID, where GWDM scored ($0.007 \pm 0.144$), substantially below DD-GAN ($0.075 \pm 0.088$) and SPAD ($0.087 \pm 0.109$), further confirming its high statistical consistency. These gains extended to patch-level metrics, where pFID dropped to 27.033, representing a 56\% improvement over SPAD (61.547), while path-FID fell to 246.881, more than 77\% below the strongest competing method, SPAD (1089.607). These large and consistent gains across all four metrics confirm that GWDM captures both global and local tissue structure more faithfully on the BCSS dataset. On MoNuSeg, GWDM achieved the best performance on the primary distributional metrics. However, the overall performance margins are comparatively modest relative to BCSS, reflecting the relatively homogeneous nature of this dataset. In terms of FID, GWDM obtained the lowest score of 45.02, surpassing the next best method, WDM (48.25), and the strongest GAN-based competitor, DD-GAN (59.09), by 6.7\% and 23.8\%, respectively. It also outperformed the weakest baseline, LDM (153.23), by a substantial margin of 70.6\%. Similarly, GWDM attained the lowest KID score of $0.020 \pm 0.001$, comparable to DD-WDM ($0.022 \pm 0.051$) and DD-GAN ($0.030 \pm 0.179$). This score is substantially lower than those of DDIM, SPAD, Imp-DDPM, and LDM, reaffirming its statistical consistency and generative stability across all datasets. At the patch level, GWDM (pFID: 51.589) performs comparably to DD-GAN (50.373), and in the path-FID case, SPAD (624.464) records a lower value than GWDM (857.123). However, these margins reflect the nature of MoNuSeg as a fundamentally low-diversity dataset. It is characterized by densely packed nuclei with highly uniform morphology, minimal staining variation, and repetitive spatial arrangement across all patches. Such structural homogeneity critically limits the extent to which any guidance-based method can differentiate itself. When all tissue regions share near-identical appearance, patch-level and trajectory-based metrics converge across methods regardless of the sophistication of the generative framework. On PAIP2019, which is a considerably large dataset, only two models were compared in order to avoid computational complexity. GWDM achieved the lowest FID of 15.76, reducing WDM's score (19.46) by 19.0\% and surpassing LDM  (209.89) by a substantial margin. The same ordering holds for KID, where GWDM  obtained the lowest score ($0.005 \pm 0.070$), ahead of WDM ($0.012 \pm 0.056$) and far below LDM ($0.254 \pm 0.001$), and for pFID, where GWDM (28.696) improved upon WDM (33.82) by 15.2\%. The only exception is Path-FID, where WDM (240.527) marginally outperforms GWDM (323.360), though both remain far ahead of LDM (1530.804). Importantly, GWDM consistently improves over WDM, the backbone on which it is built across FID, KID, and pFID, directly measuring the contribution of the proposed guidance components. Qualitative evaluation across all datasets confirmed that GWDM generates coherent, diverse, and visually realistic images, demonstrating gains in both metric-based performance and practical visual quality Fig.~\ref{fig:GWDM synthetic samples}. Overall, GWDM consistently achieved lower FID, KID and pFID scores than state-of-the-art models, proving superior realism and fidelity while maintaining competitive path-FID across datasets.

\begin{table*}[htpb]
\renewcommand{\arraystretch}{1.5}
\centering
\caption{Segmentation results on MoNuSeg and BCSS. Best values in \textbf{bold}.}
\label{tab:results}
\setlength{\tabcolsep}{5pt}
\begin{tabular}{l l l ccc ccc ccc}
\hline
\toprule

\multirow{2}{*}{\textbf{Dataset}}
  & \multirow{2}{*}{\makecell[l]{\textbf{SSL Pretraining}\\\textbf{Data}}}
  & \multirow{2}{*}{\textbf{Encoder}}
  & \multicolumn{3}{c}{\textbf{DeepLab\,v3}}
  & \multicolumn{3}{c}{\textbf{UNet++}}
  & \multicolumn{3}{c}{\textbf{SegFormer}} \\
\cmidrule(lr){4-6}
\cmidrule(lr){7-9}
\cmidrule(lr){10-12}
  & & 
  & Dice & mIoU & pAcc
  & Dice & mIoU & pAcc
  & Dice & mIoU & pAcc \\
\midrule
\multirow{3}{*}{\textbf{MoNuSeg}}
    & Real only 
    & MoCo\,v2 ResNet-50
    & 0.931  & 0.872 & 0.946
    & 0.954 & 0.914 & 0.965
    & 0.926 & 0.864 & 0.943\\
    
    &\makecell{\makecell{\text{Real} \\ \footnotesize{(~2k)}} $+$ \makecell{\text{Synthetic} \\ \footnotesize{(10k)}}}
    & MoCo\,v2 ResNet-50
    & 0.933 & 0.876 & 0.948 
    & 0.956 & 0.916 & 0.965
    & 0.928 & 0.868 & 0.945\\
    
    &\makecell{\makecell{\text{Real} \\ \footnotesize{(2k)}} $+$ \makecell{\text{Synthetic} \\ \footnotesize{(50k)}}}
    & MoCo\,v2 ResNet-50
    & \textbf{0.936} & \textbf{0.882} &  \textbf{0.951}
    & \textbf{0.956} & \textbf{0.917} & \textbf{0.966}
    & \textbf{0.930} & \textbf{0.872} & \textbf{0.946} \\
\midrule
\multirow{3}{*}{\textbf{BCSS}}
    & Real only
    & MoCo\,v2 ResNet-50
    & 0.738 & 0.599 & 0.785
    & 0.719 & 0.575 & 0.769
    & 0.727 & 0.584 & 0.776 \\

    &\makecell{\makecell{\text{Real} \\ \footnotesize{(20k)}} $+$ \makecell{\text{Synthetic} \\ \footnotesize{(10k)}}}
    & MoCo\,v2 ResNet-50
    &  0.738 & 0.598 & 0.788
    & \textbf{0.730} & 0.588 & 0.778
    &  \textbf{0.728} & 0.586 &  0.781 \\

    &\makecell{\makecell{\text{Real} \\ \footnotesize{(20k)}} $+$ \makecell{\text{Synthetic} \\ \footnotesize{(50k)}}}
    & MoCo\,v2 ResNet-50
    &  \textbf{0.745} &  \textbf{0.606} &  \textbf{0.791}
    &  0.728 & \textbf{0.586} &  \textbf{0.782} 
    &  \textbf{0.728} &  \textbf{0.586} &  \textbf{0.785} \\
\bottomrule
\end{tabular}
\label{tab:results_table}
\end{table*}
\begin{figure*}[t]
    \centering
    \includegraphics[width=1.0\textwidth]{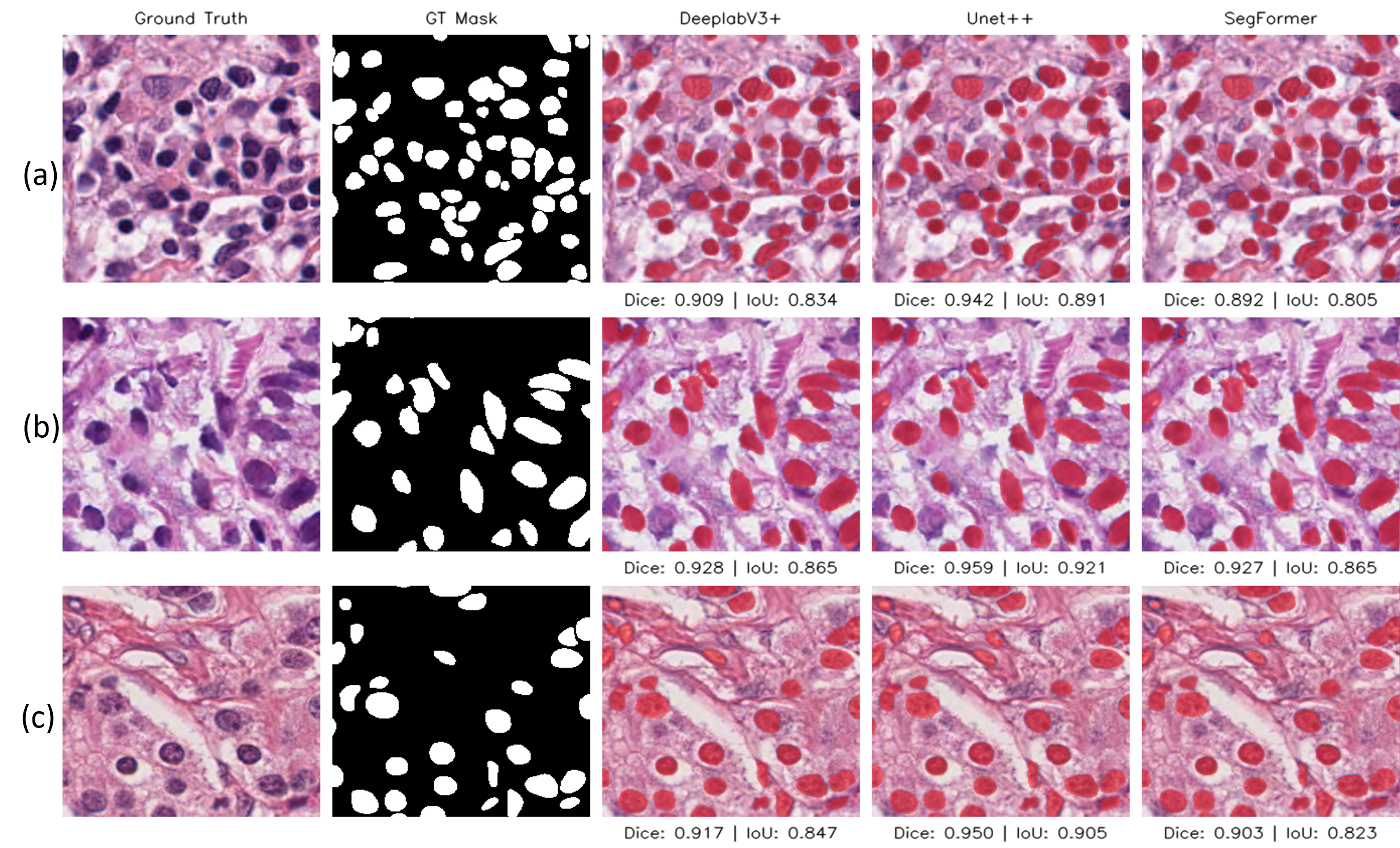}
    \caption{Segmentation results of DeepLabV3+, U-Net++, and SegFormer on the 
    MoNuSeg dataset under three training settings: (a) real data only (2k), 
    (b) 10k synthetic $+$ 2k real, and (c) 50k synthetic $+$ 2k real. 
    Downstream performance improves consistently as more synthetic data is added.}
    \label{fig:seg1}
\end{figure*}

\begin{figure*}[t]
    \centering
    \includegraphics[width=1.0\textwidth]{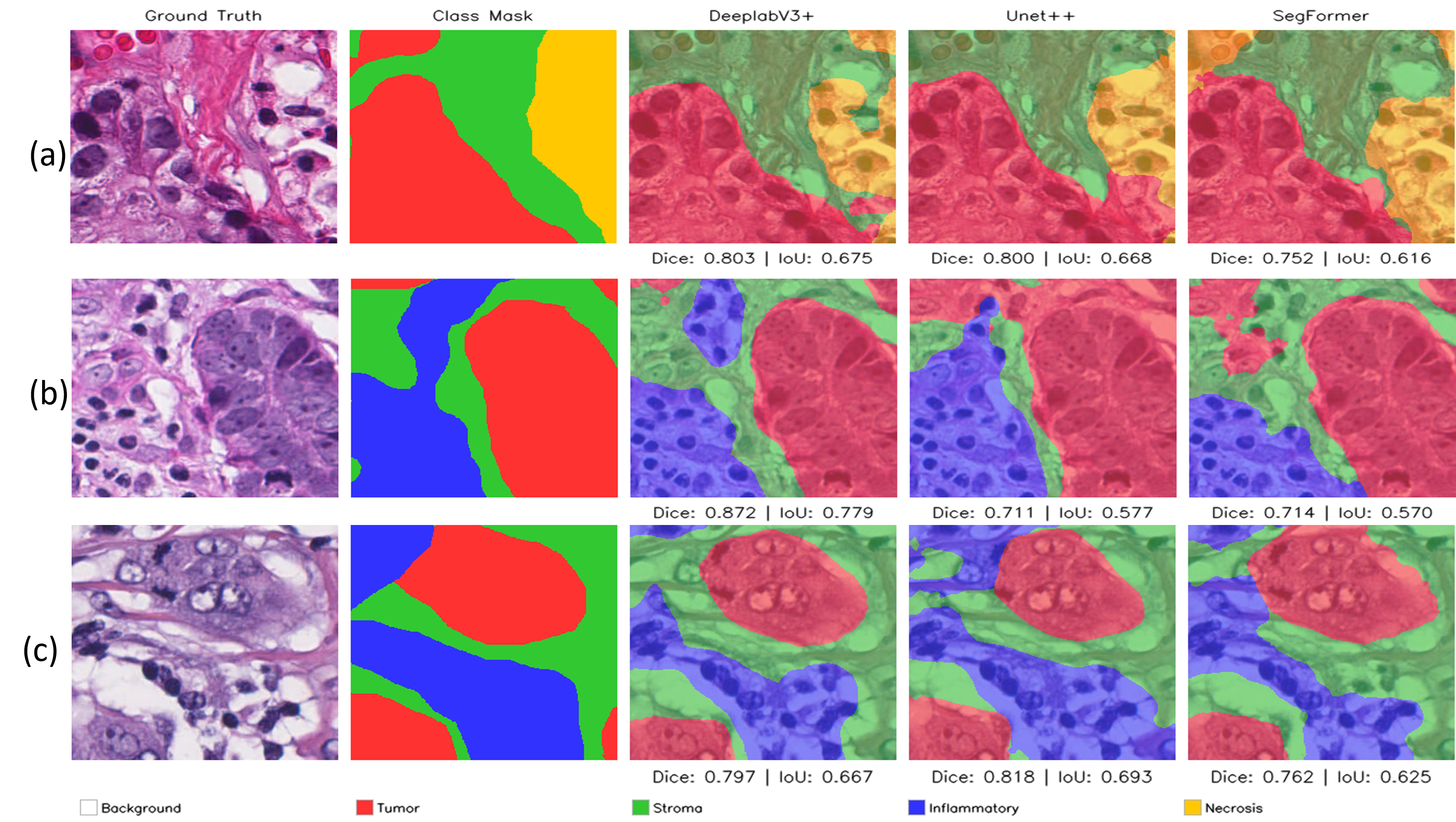}
    \caption{Multiclass segmentation results of DeepLabV3+, U-Net++, and SegFormer on the BCSS dataset under three training settings: (a) real data 
    only (2k), (b) 10k synthetic $+$ 2k real, and (c) 50k synthetic $+$ 2k real. 
    Increasing the amount of synthetic data consistently enhances downstream performance.}
    \label{fig:seg2}
\end{figure*}

\subsection{Ablation Study}
The ablation study demonstrates that each proposed guidance component effectively improves histopathology image patch synthesis on the MoNuSeg dataset. As shown in Table~\ref{tab:monuseg_ablation}, the baseline WDM, which relies solely on the wavelet reconstruction loss $\mathcal{L}_{\mathrm{rec}}$, achieves an FID of 51.49, indicating limited structural and perceptual quality in the absence of auxiliary guidance. Introducing the stain guidance term $\mathcal{L}_{\mathrm{stain}}$ alone reduces the FID to 48.80, while introducing the spatial guidance term $\mathcal{L}_{\mathrm{spat}}$ alone lowers it to 48.01. Further incorporating the morphology-aware guidance term $\mathcal{L}_{\mathrm{morph}}$ alone reduces the FID to 47.20, confirming its complementary contribution to structural fidelity. Moreover, simply combining all guidance terms with fixed weights proves insufficient; the variant without loss ramping reaches an FID of only 46.42, suggesting that an abrupt application of all constraints simultaneously can interfere with stable optimisation. Finally, when all guidance terms are combined with the proposed loss ramping strategy, the full model achieves the best overall performance, 
reducing the FID to 45.02, which corresponds to a 12.6\% reduction relative to the baseline. Overall, these progressive gains validate the proposed multi-term objective, demonstrating that stain, morphology, and spatial guidance. Together with the loss ramping strategy, jointly enhance the realism and fidelity of the generated histopathology images.

\section{Downstream Segmentation}
To evaluate the clinical utility of the proposed generative framework, we conduct downstream segmentation experiments on two benchmark histopathology datasets: MoNuSeg and BCSS. Three widely used segmentation architectures with different inductive biases are considered: DeepLabV3+ \cite{liu2024image}, UNet++ \cite{zhou2018unet++}, and SegFormer \cite{xie2021segformer}. All encoders are pretrained using MoCoV2 with a ResNet-50 backbone. Each model is trained under three settings: (i) real images only, (ii) real image combined with 10k synthetic images, and (iii) real image combined with 50k synthetic images generated by the proposed GWDM. Performance is evaluated using Dice score, mIoU, and pixel accuracy (pAcc) on held-out test sets under identical augmentation and optimisation settings.

\noindent\textbf{MoNuSeg.}
As shown in Table~\ref{tab:results_table}, incorporating GWDM generated synthetic data consistently improves segmentation performance across all three architectures. UNet++ achieves the highest overall performance, reaching a Dice of 0.956 and mIoU of 0.917 with 50k synthetic images, compared to 0.954 and 0.914 with real data only. DeepLabV3+ shows the most notable gain, improving from Dice 0.931 to 0.936 (+0.5\%) and mIoU from 0.872 to 0.882 (+1.0\%) with 50k synthetic augmentation. SegFormer similarly improves from 0.926 to 0.930 in Dice. These consistent gains across all architectures confirm that GWDM-generated synthetic data effectively enriches nuclear morphology representation, particularly beneficial in data-limited settings.

\noindent\textbf{BCSS.}
On the more challenging five-class BCSS dataset (see Table \ref{tab:results_table}), DeepLabV3+ benefits most from synthetic augmentation, improving Dice from 0.738 to 0.745 and mIoU from 0.599 to 0.606 with 50k synthetic images. UNet++ improves from Dice 0.719 to 0.730 (+1.1\%) and mIoU from 0.575 to 0.588 (+1.3\%) with 10k synthetic images, demonstrating that even limited synthetic augmentation yields meaningful gains. SegFormer shows marginal improvement, suggesting that transformer-based architectures may require larger synthetic volumes to fully benefit from augmentation. Overall, results confirm that GWDM-generated synthetic data consistently improves segmentation, particularly under data-limited conditions where annotated samples are scarce.
\begin{table*}[t]
\centering
\caption{Segmentation performance (\%) by difficulty on BCSS and MoNuSeg. Easy/Hard subsets are defined by the median real-only DeepLabV3+ per-image IoU. Best results are highlighted in bold.}
\label{tab:difficulty_both}
\setlength{\tabcolsep}{4pt}
\begin{tabular}{lllcccccc}
\toprule
& & & \multicolumn{2}{c}{Overall} & \multicolumn{2}{c}{Easy} & \multicolumn{2}{c}{Hard} \\
\cmidrule(lr){4-5}\cmidrule(lr){6-7}\cmidrule(lr){8-9}
Dataset & Model & Metric & Real & Real+Syn & Real & Real+Syn & Real & Real+Syn \\
\midrule
\multirow{6}{*}{\rotatebox{90}{BCSS}}
 & \multirow{2}{*}{DeepLabv3+} & IoU  & 53.9 & \textbf{55.2} & \textbf{69.3} & 67.2 & 38.5 & \textbf{43.2} \\
 &                             & Dice & 61.5 & \textbf{63.1} & \textbf{78.3} & 76.0 & 44.8 & \textbf{50.2} \\
 & \multirow{2}{*}{UNet++}     & IoU  & 50.1 & \textbf{55.2} & 60.5 & \textbf{67.0} & 39.7 & \textbf{43.3} \\
 &                             & Dice & 57.6 & \textbf{62.9} & 69.1 & \textbf{75.6} & 46.2 & \textbf{50.2} \\
 & \multirow{2}{*}{SegFormer}  & IoU  & 50.4 & \textbf{53.1} & 61.1 & \textbf{64.5} & 39.8 & \textbf{41.7} \\
 &                             & Dice & 58.1 & \textbf{60.8} & 69.7 & \textbf{73.0} & 46.6 & \textbf{48.6} \\
\midrule
\multirow{6}{*}{\rotatebox{90}{MoNuSeg}}
 & \multirow{2}{*}{DeepLabv3+} & IoU  & 86.8 & \textbf{87.3} & 89.4 & \textbf{89.9} & 84.2 & \textbf{84.7} \\
 &                             & Dice & 92.8 & \textbf{93.0} & 94.3 & \textbf{94.6} & 91.2 & \textbf{91.5} \\
 & \multirow{2}{*}{UNet++}     & IoU  & 91.2 & \textbf{91.4} & 93.6 & \textbf{93.9} & 88.8 & \textbf{88.9} \\
 &                             & Dice & 95.3 & \textbf{95.4} & 96.7 & \textbf{96.8} & 93.9 & \textbf{94.0} \\
 & \multirow{2}{*}{SegFormer}  & IoU  & 86.0 & \textbf{86.4} & 88.5 & \textbf{88.9} & 83.4 & \textbf{83.9} \\
 &                             & Dice & 92.3 & \textbf{92.5} & 93.8 & \textbf{94.0} & 90.8 & \textbf{91.1} \\
\bottomrule
\end{tabular}
\end{table*}

\noindent\textbf{Qualitative Results.}
Fig. \ref{fig:seg1} and Fig. \ref{fig:seg2} present qualitative segmentation results across both datasets, including the input image, ground truth mask, and predicted outputs for both binary nuclei segmentation and multiclass tissue classification (tumor, stroma, inflammatory, and necrotic regions). The figures also present per image Dice and IoU scores for each architecture. Visual inspection shows that models trained with GWDM-based synthetic augmentation produce more accurate segmentation masks, with improved boundary delineation and reduced false positives, particularly in complex tissue regions.
\section{Discussion}
Our findings show that the proposed GWDM consistently outperforms conventional generative models across multiple histopathology datasets.  It achieves state-of-the-art or competitive performance among diffusion-based methods. The generated images show high fidelity, realistic textures, and strong structural consistency  (Fig.~\ref{fig:GWDM synthetic samples}). The key distinction from standard diffusion baselines lies in the domain-informed measurement operator at the core of GWDM. This operator explicitly encodes three types of features: morphological features (nuclei density, edges, and texture), multi-scale spatial features spanning fine cellular detail to broader tissue architecture, and staining features that standardise H\&E contributions. Together, these guide the model to emphasise biologically meaningful tissue structures at each denoising step. As a result, the model avoids learning generic image statistics. This targeted conditioning improves robustness to variations in resolution and tissue composition. It also mitigates common failure modes such as mode collapse and structural inconsistency~\cite{In4, In14}, yielding more stable convergence and stronger generalization.

Quantitative results in Table~\ref{tab:table_1} confirm these observations across all three benchmarks. On BCSS, GWDM achieves the lowest FID and KID among all compared methods. This indicates that its samples most closely follow the real tissue distribution while maintaining strong diversity. On MoNuSeg, the dataset is relatively homogeneous, comprising densely packed nuclei with uniform morphology. Despite this limited scope for differentiation, GWDM remains competitive on distribution-level metrics. On PAIP2019, GWDM achieves the lowest KID, demonstrating the strongest statistical consistency with real data across cancer subtypes. These consistent trends confirm the generalization of GWDM and also validate that the domain-specific priors help the model learn biologically meaningful representations rather than superficial image statistics. Qualitative results are shown in Fig.~\ref{fig:GWDM synthetic samples}. GWDM faithfully reproduces nuclei organisation, staining balance, and tissue architecture. It also normalises colour variation effectively. Breast-cancer 
samples exhibit distinct cellular boundaries nearly indistinguishable from real images (Fig.~\ref{fig:GWDM synthetic samples}a). Liver-cancer samples show compact, well-defined structures (Fig.~\ref{fig:GWDM synthetic samples}b). MoNuSeg samples display complex cellular boundaries 
(Fig.~\ref{fig:GWDM synthetic samples}c). The clinical relevance of GWDM extends beyond generation quality. This was validated through downstream segmentation experiments. GWDM-synthesised patches were used to augment a self-supervised pretraining and limited-label fine-tuning pipeline. As reported in Table~\ref{tab:results_table}, consistent gains in mIoU, Dice, and pixel accuracy were observed across DeepLabV3+, UNet++, and SegFormer. These gains confirm that the structural fidelity of GWDM-generated images translates directly into improved segmentation generalisation. This is particularly significant in digital pathology, where large-scale annotated datasets remain costly to obtain~\cite{koohbanani2021self}. To further investigate the impact of the proposed synthetic augmentation, we stratified the test images into \textbf{Easy} and \textbf{Hard} subsets based on the median per-image IoU obtained by the real-only DeepLabV3+ baseline. The results, summarized in Table~\ref{tab:difficulty_both}, show that the proposed method consistently improves segmentation performance across both subsets. Notably, the largest gains are observed on the \textbf{Hard} subset of the BCSS dataset, indicating that the synthetic data are particularly beneficial for challenging cases with complex tissue morphology. In contrast, improvements on the \textbf{Easy} subset are smaller but remain consistent. For MoNuSeg, where the baseline performance is already close to saturation, synthetic augmentation provides modest yet uniform gains across both difficulty levels. These findings suggest that the proposed approach primarily enhances robustness on difficult samples while preserving performance on easier cases.

Several limitations remain, as the current implementation operates at the patch level and relies partly on handcrafted descriptors, which may limit scalability. Future work will explore WSI synthesis and learned pathology embeddings, including transformer-based tissue encoders and foundation models such as UNI~\cite{chen2024towards},  to improve generalisation and reduce reliance on manual feature design. Overall, GWDM offers a principled and domain-aware approach to high-fidelity histopathology image generation, providing a strong foundation for synthetic data augmentation in computational pathology workflows.

\section{Conclusion}

In this work, we introduced GWDM, a guided wavelet diffusion framework for histopathology image generation. It incorporates domain-specific tissue priors directly into the diffusion process. Specifically, GWDM integrates multiscale morphological, spatial, and staining features. These are combined with wavelet-based parameterisation and a dual perceptual-structural loss. Together, they guide the model toward biologically meaningfulreconstruction. The result is high-frequency cellular detail, stain consistency, and strong structural fidelity. Experiments were conducted across three benchmarks: BCSS, MoNuSeg, and PAIP2019. GWDM achieved consistent improvements in FID, KID, pFID and path-FID across all datasets. The generated images exhibit realistic tissue textures and nuclear organisation that closely align with true pathological appearance. Downstream segmentation experiments further validate these findings. GWDM-synthesised images consistently improved performance across DeepLabV3+, UNet++, and  SegFormer. This confirms their practical value for training robust segmentation models in clinical digital pathology settings.

\section{acknowledgments}
The authors acknowledge the support and resources that contributed to this research. This research did not receive any specific grant from funding agencies in the public, commercial, or not-for-profit sectors.
\bibliography{cas-refs}

@inproceedings{In1,
  title={Diffusion-based generation of Histopathological Whole Slide Images at a Gigapixel scale},
  author={Harb, Robert and Pock, Thomas and M{\"u}ller, Heimo},
  booktitle={Proceedings of the IEEE/CVF Winter Conference on Applications of Computer Vision},
  pages={5131--5140},
  year={2024}
}

@article{In3,
  title={Auto-encoding variational bayes},
  author={Kingma, Diederik P},
  journal={arXiv preprint arXiv:1312.6114},
  year={2013}
}

@article{In4,
  title={Generative adversarial nets},
  author={Goodfellow, Ian and Pouget-Abadie, Jean and Mirza, Mehdi and Xu, Bing and Warde-Farley, David and Ozair, Sherjil and Courville, Aaron and Bengio, Yoshua},
  journal={Advances in neural information processing systems},
  volume={27},
  year={2014}}

@article{In5,
  title={Tackling the generative learning trilemma with denoising diffusion gans},
  author={Xiao, Zhisheng and Kreis, Karsten and Vahdat, Arash},
  journal={arXiv preprint arXiv:2112.07804},
  year={2021}
}

@inproceedings{In9,
  title={A morphology focused diffusion probabilistic model for synthesis of histopathology images},
  author={Moghadam, Puria Azadi and Van Dalen, Sanne and Martin, Karina C and Lennerz, Jochen and Yip, Stephen and Farahani, Hossein and Bashashati, Ali},
  booktitle={Proceedings of the IEEE/CVF winter conference on applications of computer vision},
  pages={2000--2009},
  year={2023}
}

@inproceedings{In10,
  title={StainDiff: Transfer Stain Styles of Histology Images with Denoising Diffusion Probabilistic Models and Self-ensemble},
  author={Shen, Yiqing and Ke, Jing},
  booktitle={International Conference on Medical Image Computing and Computer-Assisted Intervention},
  pages={549--559},
  year={2023},
  organization={Springer}
}

@article{In12,
  title={Robbins basic pathology},
  author={Mitchell, Richard Sheppard and Kumar, Vinay and Abbas, Abul K and Fausto, Nelson},
  journal={Philadelphia, PA: Saunders},
  volume={8},
  pages={72--79},
  year={2007}
}

@article{In13,
  title={A Comprehensive Review of Computer-Aided Models for Breast Cancer Diagnosis Using Histopathology Images},
  author={Labrada, Alberto and Barkana, Buket D},
  journal={Bioengineering},
  volume={10},
  number={11},
  pages={1289},
  year={2023},
  publisher={MDPI}
}

@article{In14,
  title={Denoising diffusion probabilistic models},
  author={Ho, Jonathan and Jain, Ajay and Abbeel, Pieter},
  journal={Advances in neural information processing systems},
  volume={33},
  pages={6840--6851},
  year={2020}
}

@article{In16,
  title={A multimodal comparison of latent denoising diffusion probabilistic models and generative adversarial networks for medical image synthesis},
  author={M{\"u}ller-Franzes, Gustav and Niehues, Jan Moritz and Khader, Firas and Arasteh, Soroosh Tayebi and Haarburger, Christoph and Kuhl, Christiane and Wang, Tianci and Han, Tianyu and Nolte, Teresa and Nebelung, Sven and others},
  journal={Scientific Reports},
  volume={13},
  number={1},
  pages={12098},
  year={2023},
  publisher={Nature Publishing Group UK London}
}

@inproceedings{In17,
  title={Wavelet diffusion models are fast and scalable image generators},
  author={Phung, Hao and Dao, Quan and Tran, Anh},
  booktitle={Proceedings of the IEEE/CVF conference on computer vision and pattern recognition},
  pages={10199--10208},
  year={2023}
}

@article{In18,
  title={Wavelet score-based generative modeling},
  author={Guth, Florentin and Coste, Simon and De Bortoli, Valentin and Mallat, Stephane},
  journal={Advances in neural information processing systems},
  volume={35},
  pages={478--491},
  year={2022}
}

@inproceedings{In19,
  title={Wdm: 3d wavelet diffusion models for high-resolution medical image synthesis},
  author={Friedrich, Paul and Wolleb, Julia and Bieder, Florentin and Durrer, Alicia and Cattin, Philippe C},
  booktitle={MICCAI Workshop on Deep Generative Models},
  pages={11--21},
  year={2024},
  organization={Springer}
}

@article{A8,
  title={PAIP 2019: Liver cancer segmentation challenge},
  author={Kim, Yoo Jung and Jang, Hyungjoon and Lee, Kyoungbun and Park, Seongkeun and Min, Sung-Gyu and Hong, Choyeon and Park, Jeong Hwan and Lee, Kanggeun and Kim, Jisoo and Hong, Wonjae and others},
  journal={Medical image analysis},
  volume={67},
  pages={101854},
  year={2021},
  publisher={Elsevier}
}

@article{A9,
title={Structured crowdsourcing enables convolutional segmentation of histology images},
  author={Amgad, Mohamed and Elfandy, Habiba and Hussein, Hagar and Atteya, Lamees A and Elsebaie, Mai AT and Abo Elnasr, Lamia S and Sakr, Rokia A and Salem, Hazem SE and Ismail, Ahmed F and Saad, Anas M and others},
  journal={Bioinformatics},
  volume={35},
  number={18},
  pages={3461--3467},
  year={2019},
  publisher={Oxford University Press}
}

@article{a12,
  title={Gans trained by a two time-scale update rule converge to a local nash equilibrium},
  author={Heusel, Martin and Ramsauer, Hubert and Unterthiner, Thomas and Nessler, Bernhard and Hochreiter, Sepp},
  journal={Advances in neural information processing systems},
  volume={30},
  year={2017}
}

@article{a15,
  title={Diffusion models beat gans on image synthesis},
  author={Dhariwal, Prafulla and Nichol, Alexander},
  journal={Advances in neural information processing systems},
  volume={34},
  pages={8780--8794},
  year={2021}
}

@article{a19,
  title={Stage-by-stage wavelet optimization refinement diffusion model for sparse-view CT reconstruction},
  author={Xu, Kai and Lu, Shiyu and Huang, Bin and Wu, Weiwen and Liu, Qiegen},
  journal={IEEE Transactions on Medical Imaging},
  volume={43},
  number={10},
  pages={3412--3424},
  year={2024},
  publisher={IEEE}
}

@inproceedings{a20,
  title={Pathldm: Text conditioned latent diffusion model for histopathology},
  author={Yellapragada, Srikar and Graikos, Alexandros and Prasanna, Prateek and Kurc, Tahsin and Saltz, Joel and Samaras, Dimitris},
  booktitle={Proceedings of the IEEE/CVF Winter Conference on Applications of Computer Vision},
  pages={5182--5191},
  year={2024}
}

@inproceedings{a21,
  title={High-resolution image synthesis with latent diffusion models},
  author={Rombach, Robin and Blattmann, Andreas and Lorenz, Dominik and Esser, Patrick and Ommer, Bj{\"o}rn},
  booktitle={Proceedings of the IEEE/CVF conference on computer vision and pattern recognition},
  pages={10684--10695},
  year={2022}
}

@inproceedings{a22,
  title={Vit-dae: Transformer-driven diffusion autoencoder for histopathology image analysis},
  author={Xu, Xuan and Kapse, Saarthak and Gupta, Rajarsi and Prasanna, Prateek},
  booktitle={International Conference on Medical Image Computing and Computer-Assisted Intervention},
  pages={66--76},
  year={2023},
  organization={Springer}
}

@inproceedings{a23,
  title={Optimising diffusion models for histopathology image synthesis},
  author={Porter, Victoria and Gault, Richard and Craig, Stephanie and James, Jacqueline},
  booktitle={The British Machine Vision Conference},
  year={2024},
  organization={BMVC}
}

@inproceedings{a26,
  title={Demystifying MMD GANs},
  author={Binkowski, Mikolaj and Sutherland, Dougal J and Arbel, Michael and Gretton, Arthur},
  booktitle={International Conference on Learning Representations (ICLR)},
  year={2018}
}

@inproceedings{szegedy2017inception,
  title={Inception-v4, inception-resnet and the impact of residual connections on learning},
  author={Szegedy, Christian and Ioffe, Sergey and Vanhoucke, Vincent and Alemi, Alexander},
  booktitle={Proceedings of the AAAI conference on artificial intelligence},
  volume={31},
  number={1},
  year={2017}
}

@phdthesis{yin2025efficient,
  title={Efficient Generative Models for Visual Synthesis},
  author={Yin, Tianwei},
  year={2025},
  school={Massachusetts Institute of Technology}
}

@article{a27,
  title={Advancing Medical Image Segmentation: Morphology-Driven Learning with Diffusion Transformer},
  author={Kang, Sungmin and Song, Jaeha and Kim, Jihie},
  journal={arXiv preprint arXiv:2408.00347},
  year={2024}
}

@article{ruifrok2001quantification,
  title={Quantification of histochemical staining by color deconvolution},
  author={Ruifrok, Arnout C and Johnston, Dennis A and others},
  journal={Analytical and quantitative cytology and histology},
  volume={23},
  number={4},
  pages={291--299},
  year={2001},
  publisher={St. Louis, MO: Science Printers and Publishers, 1985-}
}

@article{schmitz2021multi,
  title={Multi-scale fully convolutional neural networks for histopathology image segmentation: from nuclear aberrations to the global tissue architecture},
  author={Schmitz, R{\"u}diger and Madesta, Frederic and Nielsen, Maximilian and Krause, Jenny and Steurer, Stefan and Werner, Ren{\'e} and R{\"o}sch, Thomas},
  journal={Medical image analysis},
  volume={70},
  pages={101996},
  year={2021},
  publisher={Elsevier}
}

@inproceedings{mescheder2018training,
  title={Which training methods for GANs do actually converge?},
  author={Mescheder, Lars and Geiger, Andreas and Nowozin, Sebastian},
  booktitle={International conference on machine learning},
  pages={3481--3490},
  year={2018},
  organization={PMLR}
}

@inproceedings{akrout2023diffusion,
  title={Diffusion-based data augmentation for skin disease classification: Impact across original medical datasets to fully synthetic images},
  author={Akrout, Mohamed and Gyepesi, B{\'a}lint and Holl{\'o}, P{\'e}ter and Po{\'o}r, Adrienn and Kincs{\H{o}}, Bl{\'a}ga and Solis, Stephen and Cirone, Katrina and Kawahara, Jeremy and Slade, Dekker and Abid, Latif and others},
  booktitle={International conference on medical image computing and computer-assisted intervention},
  pages={99--109},
  year={2023},
  organization={Springer}
}

@article{li2023ddmm,
  title={Ddmm-synth: A denoising diffusion model for cross-modal medical image synthesis with sparse-view measurement embedding},
  author={Li, Xiaoyue and Shang, Kai and Wang, Gaoang and Butala, Mark D},
  journal={arXiv preprint arXiv:2303.15770},
  year={2023}
}

@inproceedings{farooq2024derm,
  title={Derm-t2im: Harnessing synthetic skin lesion data via stable diffusion models for enhanced skin disease classification using vit and cnn},
  author={Farooq, Muhammad Ali and Yao, Wang and Schukat, Michael and Little, Mark A and Corcoran, Peter},
  booktitle={2024 46th Annual International Conference of the IEEE Engineering in Medicine and Biology Society (EMBC)},
  pages={1--5},
  year={2024},
  organization={IEEE}
}

@article{chen2024towards,
  title={Towards a general-purpose foundation model for computational pathology},
  author={Chen, Richard J and Ding, Tong and Lu, Ming Y and Williamson, Drew FK and Jaume, Guillaume and Song, Andrew H and Chen, Bowen and Zhang, Andrew and Shao, Daniel and Shaban, Muhammad and others},
  journal={Nature medicine},
  volume={30},
  number={3},
  pages={850--862},
  year={2024},
  publisher={Nature Publishing Group US New York}
}

@inproceedings{konz2024anatomically,
  title={Anatomically-controllable medical image generation with segmentation-guided diffusion models},
  author={Konz, Nicholas and Chen, Yuwen and Dong, Haoyu and Mazurowski, Maciej A},
  booktitle={International Conference on Medical Image Computing and Computer-Assisted Intervention},
  pages={88--98},
  year={2024},
  organization={Springer}
}

@article{chambon2022roentgen,
  title={Roentgen: vision-language foundation model for chest x-ray generation},
  author={Chambon, Pierre and Bluethgen, Christian and Delbrouck, Jean-Benoit and Van der Sluijs, Rogier and Po{\l}acin, Ma{\l}gorzata and Chaves, Juan Manuel Zambrano and Abraham, Tanishq Mathew and Purohit, Shivanshu and Langlotz, Curtis P and Chaudhari, Akshay},
  journal={arXiv preprint arXiv:2211.12737},
  year={2022}
}

@article{ho2022classifier,
  title={Classifier-free diffusion guidance},
  author={Ho, Jonathan and Salimans, Tim},
  journal={arXiv preprint arXiv:2207.12598},
  year={2022}
}

@inproceedings{zhang2023adding,
  title={Adding conditional control to text-to-image diffusion models},
  author={Zhang, Lvmin and Rao, Anyi and Agrawala, Maneesh},
  booktitle={Proceedings of the IEEE/CVF international conference on computer vision},
  pages={3836--3847},
  year={2023}
}

@inproceedings{vaishnani2023impact,
  title={Impact of stain normalisation technique on deep learning based nuclei segmentation in histopathological image},
  author={Vaishnani, Kishankumar and Gohel, Bakul and Hati, Avik},
  booktitle={2023 International Conference on Advances in Intelligent Computing and Applications (AICAPS)},
  pages={1--4},
  year={2023},
  organization={IEEE}
}

@article{rani2025efficient,
  title={Efficient real-world image denoising using multi-scale gaussian pyramids},
  author={Rani, Asha and Bhogal, Rosepreet Kaur},
  journal={Scientific Reports},
  volume={15},
  number={1},
  pages={40086},
  year={2025},
  publisher={Nature Publishing Group UK London}
}

@inproceedings{moghadam2023morphology,
  title={A morphology focused diffusion probabilistic model for synthesis of histopathology images},
  author={Moghadam, Puria Azadi and Van Dalen, Sanne and Martin, Karina C and Lennerz, Jochen and Yip, Stephen and Farahani, Hossein and Bashashati, Ali},
  booktitle={Proceedings of the IEEE/CVF winter conference on applications of computer vision},
  pages={2000--2009},
  year={2023}
}

@inproceedings{shrivastava2023nasdm,
  title={Nasdm: Nuclei-aware semantic histopathology image generation using diffusion models},
  author={Shrivastava, Aman and Fletcher, P Thomas},
  booktitle={international conference on medical image computing and computer-assisted intervention},
  pages={786--796},
  year={2023},
  organization={Springer}
}

@article{kazerouni2022diffusion,
  title={Diffusion models for medical image analysis: A comprehensive survey},
  author={Kazerouni, Amirhossein and Aghdam, Ehsan Khodapanah and Heidari, Moein and Azad, Reza and Fayyaz, Mohsen and Hacihaliloglu, Ilker and Merhof, Dorit},
  journal={arXiv preprint arXiv:2211.07804},
  year={2022}
}

@article{ozbey2023unsupervised,
  title={Unsupervised medical image translation with adversarial diffusion models},
  author={{\"O}zbey, Muzaffer and Dalmaz, Onat and Dar, Salman UH and Bedel, Hasan A and {\"O}zturk, {\c{S}}aban and G{\"u}ng{\"o}r, Alper and Cukur, Tolga},
  journal={IEEE Transactions on Medical Imaging},
  volume={42},
  number={12},
  pages={3524--3539},
  year={2023},
  publisher={IEEE}
}

@inproceedings{lee2025lesion,
  title={Lesion-Aware Post-training of Latent Diffusion Models for Synthesizing Diffusion MRI from CT Perfusion},
  author={Lee, Junhyeok and Kim, Hyunwoong and Chung, Hyungjin and Eom, Heeseong and Jang, Joon and Sohn, Chul-Ho and Choi, Kyu Sung},
  booktitle={International Conference on Medical Image Computing and Computer-Assisted Intervention},
  pages={282--291},
  year={2025},
  organization={Springer}
}

@inproceedings{bhosale2025pathdiff,
  title={Pathdiff: Histopathology image synthesis with unpaired text and mask conditions},
  author={Bhosale, Mahesh and Wasi, Abdul and Zhai, Yuanhao and Tian, Yunjie and Border, Samuel and Xi, Nan and Sarder, Pinaki and Yuan, Junsong and Doermann, David and Gong, Xuan},
  booktitle={Proceedings of the IEEE/CVF International Conference on Computer Vision},
  pages={22415--22424},
  year={2025}
}

@article{kumar2019multi,
  title={A multi-organ nucleus segmentation challenge},
  author={Kumar, Neeraj and Verma, Ruchika and Anand, Deepak and Zhou, Yanning and Onder, Omer Fahri and Tsougenis, Efstratios and Chen, Hao and Heng, Pheng-Ann and Li, Jiahui and Hu, Zhiqiang and others},
  journal={IEEE transactions on medical imaging},
  volume={39},
  number={5},
  pages={1380--1391},
  year={2019},
  publisher={IEEE}
}

@incollection{wang2016practical,
  title={A practical guide to the cancer genome atlas (TCGA)},
  author={Wang, Zhining and Jensen, Mark A and Zenklusen, Jean Claude},
  booktitle={Statistical genomics: methods and protocols},
  pages={111--141},
  year={2016},
  publisher={Springer}
}

@inproceedings{park2019semantic,
  title={Semantic image synthesis with spatially-adaptive normalization},
  author={Park, Taesung and Liu, Ming-Yu and Wang, Ting-Chun and Zhu, Jun-Yan},
  booktitle={Proceedings of the IEEE/CVF conference on computer vision and pattern recognition},
  pages={2337--2346},
  year={2019}
}

@inproceedings{nichol2021improved,
  title={Improved denoising diffusion probabilistic models},
  author={Nichol, Alexander Quinn and Dhariwal, Prafulla},
  booktitle={International conference on machine learning},
  pages={8162--8171},
  year={2021},
  organization={PMLR}
}

@article{song2020denoising,
  title={Denoising diffusion implicit models},
  author={Song, Jiaming and Meng, Chenlin and Ermon, Stefano},
  journal={arXiv preprint arXiv:2010.02502},
  year={2020}
}

@article{liu2024image,
  title={Image semantic segmentation approach based on DeepLabV3 plus network with an attention mechanism},
  author={Liu, Yanyan and Bai, Xiaotian and Wang, Jiafei and Li, Guoning and Li, Jin and Lv, Zengming},
  journal={Engineering Applications of Artificial Intelligence},
  volume={127},
  pages={107260},
  year={2024},
  publisher={Elsevier}
}

@inproceedings{zhou2018unet++,
  title={Unet++: A nested u-net architecture for medical image segmentation},
  author={Zhou, Zongwei and Rahman Siddiquee, Md Mahfuzur and Tajbakhsh, Nima and Liang, Jianming},
  booktitle={International workshop on deep learning in medical image analysis},
  pages={3--11},
  year={2018},
  organization={Springer}
}

@article{xie2021segformer,
  title={SegFormer: Simple and efficient design for semantic segmentation with transformers},
  author={Xie, Enze and Wang, Wenhai and Yu, Zhiding and Anandkumar, Anima and Alvarez, Jose M and Luo, Ping},
  journal={Advances in neural information processing systems},
  volume={34},
  pages={12077--12090},
  year={2021}
}

@article{koohbanani2021self,
  title={Self-path: Self-supervision for classification of pathology images with limited annotations},
  author={Koohbanani, Navid Alemi and Unnikrishnan, Balagopal and Khurram, Syed Ali and Krishnaswamy, Pavitra and Rajpoot, Nasir},
  journal={IEEE Transactions on Medical Imaging},
  volume={40},
  number={10},
  pages={2845--2856},
  year={2021},
  publisher={IEEE}
}

@article{aydin2026improving,
  title={Improving histopathological image classification with patch-based embeddings and ensemble learning: a study on enteroscope biopsy images},
  author={Aydin, Musa},
  journal={International Journal of Machine Learning and Cybernetics},
  volume={17},
  number={3},
  pages={120},
  year={2026},
  publisher={Springer}
}
\bibliographystyle{unsrt}
\end{document}